\documentclass[prl,showpacs,superscriptaddress]{revtex4}
\usepackage{graphicx,epsfig}
\usepackage{times}
\usepackage{graphics,dcolumn,bm,float}
\usepackage{amssymb,amsmath,rotate,color}

\begin{document}
\unitlength 1 cm
\newtheorem{thm}{Theorem}
\newtheorem{lem}[thm]{Lemma}

\newcommand{\be}{\begin{equation}}
\newcommand{\ee}{\end{equation}}
\newcommand{\bearr}{\begin{eqnarray}}
\newcommand{\eearr}{\end{eqnarray}}
\newcommand{\nn}{\nonumber}
\newcommand{\vk}{\vec k}
\newcommand{\vp}{\vec p}
\newcommand{\vq}{\vec q}
\newcommand{\vkp}{\vec {k'}}
\newcommand{\vpp}{\vec {p'}}
\newcommand{\vqp}{\vec {q'}}
\newcommand{\bk}{{\bf k}}
\newcommand{\bp}{{\bf p}}
\newcommand{\bq}{{\bf q}}
\newcommand{\br}{{\bf r}}
\newcommand{\up}{\uparrow}
\newcommand{\down}{\downarrow}
\newcommand{\fns}{\footnotesize}
\newcommand{\ns}{\normalsize}
\newcommand{\cdag}{c^{\dagger}}

\title{Analysis the neutral triplet collective mode in honeycomb lattice: without one-cone approximation}
\author{M. Ebrahimkhas}
\address{Department of physics, Mahabad Branch, Islamic Azad University, Mahabad 59135, Iran}
\begin{abstract}
  In this research has been studied the possibility of existence   the neutral triplet collective mode
  in undoped graphene  and graphite without one-cone approximation. This work tries to study this collective mode  from different points of view. It also studies the effects of
  on-site Hubbard  interaction on neutral collective mode in honeycomb lattice via two representations (sub-lattice and eigenvector).
  The effective long rang interaction in honeycomb lattice has been introduced for studying  the spin collective mode. This work shows that short and effective long range
  interactions  are not responsible   for the formation of neutral   triplet collective mode in graphene without {\it one-cone} approximation.
\end{abstract}


\maketitle
\section{ Introduction}
The discovery of graphene, one layer of carbon atoms, arose high hopes for its application in future of
electronic devices~\cite{Neto-RMP}. One reason for this attention is high  mobility of charge carriers (electrons or holes)
whose density are well controlled  by electrical and chemical doping. 
Although the application of pure graphene in electronic industry didn't develop because of  the gap-less dispersion.
Adding atoms and doping in graphene $e. g.$ fluorinated mono-vacancies~\cite{Kaloni-EPL-100},
Ge-interrelated~\cite{Kaloni-EPL-99} have been studied for investigation the electrical 
characteristics of graphene under impurity doping, which  gives rise to exploring new interesting electronic and magnetic properties.  

 The prediction of existence the neutral triplet collective
mode in a honeycomb lattice is considered to be the new idea~\cite{Baskaran-Jafari}.
This collective mode was predicted by considering random phase approximation (RPA) method in Hubbard model on 
the honeycomb lattice $e. g.$ undoped graphene and graphite. By considering this collective mode,
good agreement between experimental and theoretical results in TRPES experiment on graphite by Momentum Average (MA) 
method was achieved~\cite{Ebrahimkhas-PRB}. In this approximation, neutral triplet spin-1 mode was considered as coupling
between two spinon bound states in undoped graphene but in scalar form not in $2\times2$ 
matrix~\cite{Jafari-Baskaran},~\cite{Baskaran-Jafari}. 
If we study and recheck the calculations of ref.~\cite{{Jafari-Baskaran}, {Ebrahimkhas-IJMPB}},
we will find, first: some parts of the calculations between two sub-lattice
are eliminated and second: effects of overlap factors  are neglected~\cite{{Peres-comment},{Jafari-recomment}}.
The innovation of ref.~\cite{Jafari-Baskaran} was considering scalar form for  spin susceptibility, instead of $2\times 2$ tensor
for two sub-lattice material.
The essential purpose of this research is analyzing the validity of spin-1 collective mode without one-cone approximation
in Hubbard model and its novelty  is calculating  the effects of effective long range 
interaction $(U^{eff}_{l-r})$ in RPA method on  neutral triplet collective mode~\cite{JPHCM.Jafari}.
Why  we considered this form for long range interaction?
Because by considering only this interaction type
we will obtain  Eq.~\ref{main.eq} which is an essential equation in finding neutral collective mode in RPA formulas
(without one-cone approximation), in addition we should consider those types of interaction,
acting between electrons with opposite spins.   
The effects of  Coulomb interaction is known to be in the form $e^{2}/(\epsilon r)$ which lead to prediction
of plasmon dispersion in graphene and has been discussed in details in ref.~\cite{Wunsch}.
The coulomb interaction in the honey comb lattice didn't result spin-1 collective mode.

 In the first section of this paper, the effects of short range Hubbard interaction in RPA method on honeycomb
lattice was calculated exactly without neglecting any interaction terms~\cite{Peres-comment}, in two
representations. In the second section, we first introduced new effective long range
interaction as a new model in graphene and then we used this model for studying the neutral triplet collective
mode. finally the finding of these calculations are summarized and discussed on validation of neutral collective mode
in undoped graphene.

\section{ On-site interaction: sub lattice representation}

  The Hubbard Hamiltonian consist of two terms: tight binding (TB) and short range Hubbard interaction (HI),
 
 \bearr
 H&=&H_{TB}+H_{HI} 
 =-t\sum_{<i,j>,\sigma}(a_{i}^{\dagger \sigma}b_{j}^{\sigma}+b_{j}^{\dagger \sigma}a_{i}^{\sigma})\nn\\
 &+& U\sum_{i}(a_{i}^{\dagger \uparrow}a_{i}^{\uparrow}a_{i}^{\dagger \downarrow}a_{i}^{\downarrow}+
 b_{i}^{\dagger \uparrow}b_{i}^{\uparrow}b_{i}^{\dagger \downarrow}b_{i}^{\downarrow}),\\
 \label{TB.eq}\nn
 \eearr
 
 where $t$ is the nearest neighbor hopping. $a^{\sigma}_{i},b^{\sigma}_{i},(a^{\dagger \sigma}_{i},b^{\dagger \sigma}_{i})$
 are annihilation (creation) operators for electrons in  sub lattices $A, B$ respectively, and $\sigma$ stands for spin of electrons,
 and $U$ is the Hubbard on-site interaction potential. The Hubbard interaction term can be written in the exchange channel,
 
 \be
 H_{HI}= -U\sum_{i}(a_{i}^{\dagger \uparrow}a_{i}^{\downarrow}a_{i}^{\dagger \downarrow}a_{i}^{\uparrow}+
 b_{i}^{\dagger \uparrow}b_{i}^{\downarrow}b_{i}^{\dagger \downarrow}b_{i}^{\uparrow})
 \label{exHI.eq}
 \ee

\begin{figure}[h]
\begin{center}
\vspace{-3mm}
\includegraphics[width=8cm,height=3cm,angle=0]{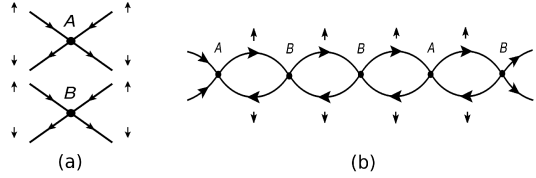}
\vspace{-3mm}
\caption{(a) Each vertex includes creation(destruction) of an electron with up (down) spin. These vertices show
interaction between two electrons with opposite spin in same atoms in sub-lattices by short range repulsive potential  $U$. This type of
interaction is shown in Eq.~\ref{exHI.eq}, in each sub-lattice $A(B)$ electrons with up(down) spin interact to electrons with down(up) spin.
(b) The RPA chain includes summing of $A$, $B$ vertices. This figure is one of the $RPA$ chains.}
 \label{vertices.fig} 
\end{center}
\vspace{-3mm}
   \end{figure}

  The  interaction term is represented by Feynman diagrams with two kinds of vertices.
 Each vertex corresponds to   the on-site repulsion interaction on the $A$, $B$ sub-lattices Fig.~\ref{vertices.fig}.
 In each vertex electrons with opposite spins interact via repulsion potential~\cite{BF}.
 Fig.(\ref{vertices.fig}-a) shows creation (annihilation) of electrons with up and down spin in each vertex
  either in $A$ or $B$ sub-lattice. In Fig.(~\ref{vertices.fig}-b), by adding vertices of 
 each sub-lattice, RPA chain is formed. In this chain the vertex with the same or different sub-lattice put to gather,
 so we  define polarization operators, $\chi_{AA}, \chi_{AB}$ which include effects of 
 overlap factors.

 Each vertex in the chain can be either of $A$ or $B$ type. By summing the RPA series, one can achieve to the
 collection of equations for susceptibilities in the sub-lattice representation~\cite{Peres-comment},
 
 \be
\begin{cases}
 \chi_{AA}=\chi^{0}_{AA}+U\chi^{0}_{AA}\chi_{AA}+U\chi^{0}_{AB}\chi_{BA}\\
 \chi_{AB}=\chi^{0}_{AB}+U\chi^{0}_{AA}\chi_{AB}+U\chi^{0}_{AB}\chi_{BB}\\
 \chi_{BA}=\chi^{0}_{BA}+U\chi^{0}_{BA}\chi_{AA}+U\chi^{0}_{BB}\chi_{BA}\\
 \chi_{BB}=\chi^{0}_{BB}+U\chi^{0}_{BA}\chi_{AB}+U\chi^{0}_{BB}\chi_{BB}\\
\end{cases}.
\label{coupling.eq}
\ee

 The coupling equations in Eq.~\ref{coupling.eq} should be decoupled, so giving two subsystems with 
 the common determinant, which should be zero at the plasmon dispersion~\cite{Peres-comment}:
 
 \be
 \begin{vmatrix} 
 1-U\chi^{0}_{AA} &    -U\chi^{0}_{AB}\\ 
  -U\chi^{0}_{BA}  &   1-U\chi^{0}_{BB}
  \end{vmatrix}=0,
  \label{determin1.eq}
\ee
 
 or 
 
 \be
 1-U(\chi^{0}_{AA}+\chi^{0}_{BB})+U^{2}\chi^{0}_{AA}\chi^{0}_{BB}-U^2 \chi^{0}_{AB}\chi^{0}_{BA}=0.
 \label{determin2.eq}
 \ee

  The bare polarization operators or susceptibilities of vertices can be calculated using non-interacting
  Green's function for graphene's electrons in the sub-lattice representation. By considering the Green's functions
  in the momentum representation and close to Dirac points, classification of Green's function
  by electron valleys $\vec{K}$ and $-\vec{K}$, by vicinities to the corresponding Dirac points was achieved. In the valleys
  $\vec{K}$ and $-\vec{K}$ the non-interacting Hamiltonians in the representation of sub lattice $A, B$ are

  \be
 H_{\pm \vec{K}}=\begin{pmatrix} 
   0          &    \pm p_{x}-ip_{y}\\ 
  \pm p_{x}+ip_{y} &   0
  \end{pmatrix},
  \label{nonintH.eq}
\ee

  here by considering $v_{F}=1$, the corresponding Matsubara Green's functions for the undoped graphene ($\mu=0$) are

\be
 G^{\pm \vec{K}}(\vec{p},i\epsilon)=\frac{1}{(i\epsilon)^2-p^2}\begin{pmatrix} 
   i\epsilon         &    \pm pe^{\mp i\phi}\\ 
  \pm pe^{\pm i\phi} &       i\epsilon
  \end{pmatrix},
  \label{MG.eq}
\ee

 where $\phi$ is the azimuthal angle of the momentum vector $\vec{p}$.
By using Eq.~\ref{MG.eq}, the polarization operators can be calculated as

\be
\chi^{0}_{\alpha,\beta}(\vec{q},i\omega)=-T\sum_{v=\pm \vec{K}}\sum_{\epsilon} \int \frac{d\vec{p}}{(2\pi)^2}
G^{v}_{\alpha, \beta}(\vec{p},i\epsilon)G^{-v}_{\beta, \alpha}(\vec{p'},i\epsilon+i\omega),
\label{chi-RPA.eq}
\ee

 where $\vec{p'}=\vec{p}+\vec{q}$. By performing the summation over fermionic Matsubara frequencies
 $\epsilon=\pi T(2n+1)$ and electron valleys $v=\pm\vec{K}$,
 
 \bearr
 \chi^{0}_{AA}(\vec{q}, i\omega)=\chi^{0}_{BB}(\vec{q}, i\omega)=-\int{\frac{d\vec{p}}{(2\pi)^2}\frac{p+p'}{(i\omega)^2-(p+p')^2}},\nn\\
  \chi^{0}_{AB}(\vec{q}, i\omega)=\chi^{0}_{BA}(\vec{q}, i\omega)=
  \int{\frac{d\vec{p}}{(2\pi)^2}\frac{cos(\phi-\phi')(p+p')}{(i\omega)^2-(p+p')^2}}.\nn\\
  \label{chi-sublattices.eq}
 \eearr 
 
   Our results in  Eq.~\ref{determin1.eq} and Eq.~\ref{chi-sublattices.eq} are in agreement with results of ref.~\cite{Peres-comment}
   , the only difference is that the decomposition of the overlap factor $cos(\phi-\phi')$ near the Dirac point was performed in 
   Eq.~\ref{chi-sublattices.eq}. Now by using Eq.~\ref{determin1.eq} and Eq.~\ref{chi-sublattices.eq} we should find zeros of 
   Eq.~\ref{determin2.eq}, but we didn't obtain  zeros or triplet collective excitation for reasonable range of $U$. In other words,
   the short-range interaction dose not lead to any neutral collective mode. 
   
\section{On-site interaction: eigenvector representation}

 In this section the same problem of RPA in the triplet channel of the on-site interaction will be considered~\cite{note},
 but in the representation of the eigenvectors of the Hamiltonians Eq.~\ref{nonintH.eq}. For the case
 of electron valleys  $\vec{K}$ and $-\vec{K}$, the eigenvectors are
 
 \be
 f^{\vec{K}}_{s} = \frac{1}{\sqrt{2}} \begin{pmatrix} 
  e^{-i\phi/2}
  \\ se^{i\phi/2}
   \end{pmatrix}, f^{-\vec{K}}_{s} = \frac{1}{\sqrt{2}} \begin{pmatrix} -se^{i\phi/2}
  \\   e^{i\phi/2} \end{pmatrix} ,
  \label{eignvector2.eq}
\ee

  where $f^{\pm \vec{K}}$ corresponds  to electron states  in the conduction band with the energy $v_{F}p$ and in the valance band
 with the energy $-v_{F}p$. The interaction Hamiltonian Eq.~\ref{exHI.eq} in the momentum representation takes
 this form:
 
 \bearr
 H_{int}=-\frac{U}{N}\sum_{\vec{p}_{1} \vec{p}_{2} \vec{q}} \left\{ \Psi^{\uparrow \dagger}_{\vec{p'}_{1}}
 \begin{pmatrix} 1  &  0  \\ 0  &  0 \end{pmatrix} \Psi^{\downarrow}_{\vec{p}_{1}} \Psi^{\downarrow \dagger}_{\vec{p'}_{2}}
    \begin{pmatrix} 1  & 0 \\ 0  & 0  \end{pmatrix}  \Psi^{\uparrow}_{\vec{p}_{2}}  \right. \nn\\
   \left. + \Psi^{\uparrow \dagger}_{\vec{p'}_{1}}
 \begin{pmatrix} 0  &  0  \\ 0  &  1 \end{pmatrix} \Psi^{\downarrow}_{\vec{p}_{1}} \Psi^{\downarrow \dagger}_{\vec{p'}_{2}}
    \begin{pmatrix} 0  & 0 \\ 0  & 1  \end{pmatrix}  \Psi^{\uparrow}_{\vec{p}_{2}}   \right \},
    \label{sum_psi1.eq}
 \eearr
   
    where $\vec{p'}_{1}=\vec{p}_{1}+\vec{q}$, $\vec{p'}_{2}=\vec{p}_{2}-\vec{q}$, and 
 $\Psi^{\uparrow \downarrow}_{\vec{p}}=(a^{\uparrow \downarrow}_{\vec{p}}, b^{\uparrow \downarrow}_{\vec{p}})^T$
 is a two-components column matrix, was made from electron destruction operators. The sum in Eq.~\ref{sum_psi1.eq} over
 $\vec{p}_{1}, \vec{p}_{2}, \vec{q}$ should be distributed among electron valleys. Retaining only the terms corresponding
 to small $\vec{q}$ (since we search small-momentum collective excitation), we get two intra-valley and two inter-valley terms

\begin{figure}[h]
\begin{center}
\vspace{-3mm}
\includegraphics[width=6cm,height=2cm,angle=0]{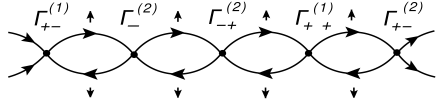}
\vspace{-3mm}
\caption{ Each vertex is including $\Gamma^{(1)}$ or $\Gamma^{(2)}$. Each vertex is include interaction between
electrons of valance or conduction in each valley. In this chain I  considered many types of $\Gamma$ functions
which described interaction of electrons in each valley. This chain is one of the $RPA$ chains.}
   \label{vertices2.fig}
\end{center}
\vspace{-3mm}
\end{figure}

\bearr
 H_{int}=-\frac{U}{N}\sum_{\vec{p}_{1} \vec{p}_{2} \vec{q}} \sum_{\vec{v}_{1},\vec{v}_{2}}
 \left\{ \Psi^{\uparrow \dagger}_{\vec{v'}_{1}}
 \begin{pmatrix} 1  &  0  \\ 0  &  0 \end{pmatrix} \Psi^{\downarrow}_{\vec{v}_{1}} \Psi^{\downarrow \dagger}_{\vec{v'}_{2}}
    \begin{pmatrix} 1  & 0 \\ 0  & 0  \end{pmatrix}  \Psi^{\uparrow}_{\vec{v}_{2}}  \right. \nn\\
   \left. + \Psi^{\uparrow \dagger}_{\vec{v'}_{1}}
 \begin{pmatrix} 0  &  0  \\ 0  &  1 \end{pmatrix} \Psi^{\downarrow}_{\vec{v}_{1}} \Psi^{\downarrow \dagger}_{\vec{v'}_{2}}
    \begin{pmatrix} 0  & 0 \\ 0  & 1  \end{pmatrix}  \Psi^{\uparrow}_{\vec{v}_{2}}   \right \},
    \label{sum_psi2.eq}
 \eearr

   where $\vec{v}_{\alpha}=\vec{V}_{\alpha}+\vec{p}_{\alpha}, \vec{v'}_{\alpha}=\vec{V}_{\alpha}+\vec{p'}_{\alpha}$ and 
   $\vec{V}_{\alpha}=\pm \vec{K}$. Now by changing the basis of the eigenvectors, the Eq.~\ref{eignvector2.eq}
   can be performed according to the relation $\Psi^{\uparrow \downarrow}_{\vec{V}+\vec{p}}=
   \sum_{s=\pm}f^{\vec{V}}_{s}c^{\uparrow \downarrow}_{\vec{V}+\vec{p},s}$, where $c^{\uparrow \downarrow}_{\vec{V}+\vec{p},s}$ is the
   destruction operator for electron from the valley $\vec{V}=\pm \vec{K}$, from the conduction $(s=+1)$ or
   valance $(s=-1)$ band, with the momentum $\vec{p}$ (measured from the corresponding Dirac point) and
   with up/down spin. The result for such transformation of Eq.~\ref{sum_psi2.eq} is,
   
   \bearr
   H_{int}=\sum_{\vec{p}_{1}\vec{p}_{2}\vec{q}}\sum_{\vec{V}_{1}\vec{V}_{2}=\pm\vec{K}}\sum_{s'_1,s'_2,s_1,s_2=\pm} \left \{
   \Gamma^{(1)}_{\vec{V}_{1}\vec{V}_{2}}+\Gamma^{(2)}_{\vec{V}_{1}\vec{V}_{2}}  \right \} \nn \\
   \times c^{\uparrow \dagger}_{\vec{V}_{1}+\vec{p'}_{1},s'_1} c^{\downarrow }_{\vec{V}_{1}+\vec{p}_{1},s_1}
   c^{\downarrow  \dagger}_{\vec{V}_{2}+\vec{p'}_{2},s'_2} c^{\uparrow }_{\vec{V}_{2}+\vec{p'}_{2},s_2}
   \label{H_int_new.eq}
   \eearr
   
   where the vertices are,
   
   \bearr
   \Gamma^{(1)}_{++}&=&\Gamma^{(2)}_{--}=\frac{u}{4}e^{\frac{i}{2}(-\phi_{1}+\phi'_{1}-\phi_{2}+\phi'_{2})},\nn \\
   \Gamma^{(2)}_{++}&=&\Gamma^{(1)}_{--}=\frac{u}{4}s_{1}s'_{1}s_{2}s'_{2}e^{\frac{i}{2}(\phi_{1}-\phi'_{1}+\phi_{2}-\phi'_{2})}, \nn\\
   \Gamma^{(1)}_{+-}&=&\Gamma^{(2)}_{-+}=\frac{u}{4}s_{2}s'_{2}e^{\frac{i}{2}(-\phi_{1}+\phi'_{1}+\phi_{2}-\phi'_{2})},\nn \\
   \Gamma^{(2)}_{+-}&=&\Gamma^{(1)}_{-+}=\frac{u}{4}s_{1}s'_{1}e^{\frac{i}{2}(\phi_{1}-\phi'_{1}-\phi_{2}+\phi'_{2})}
   \label{vertices4.eq}
   \eearr

  where $u=U/N$. The interaction Hamiltonian Eq.~\ref{H_int_new.eq} gives rise to RPA series of the type
  depicted in Fig.~\ref{vertices2.fig}. A vertex in each place of these series can be of type $\Gamma^{(1,2)}$,
  while electron valleys in the Green function between vertices can be $\vec{K}$ or $-\vec{K}$. 
  
  The full vertex can be renormalized by RPA summation, $\Gamma^{(i,j)}_{\vec{V}_{1}\vec{V}_{2}}$, where $i,j=1,2$ are the types of
  left and right vertices in the series. The Bethe-Sal peter equations for this vertex is,
  
  \bearr
 &&\Gamma^{(ij)}_{\vec{V}_{1}\vec{V}_{2}}(\vec{p'}_{1},s'_{1},\vec{p}_{1},s_{1},\vec{p'}_{2},s'_{2}, \vec{p}_{2},s_{2};i\omega)= \nn\\
  &&\delta_{ij}\Gamma^{(i)}_{\vec{V}_{1}\vec{V}_{2}}(\vec{p'}_{1},s'_{1},\vec{p}_{1},s_{1},\vec{p'}_{2},s'_{2}, \vec{p}_{2},s_{2}) \nn \\
  &&-T\sum_{k\vec{V}}\sum_{\epsilon ss'}\int \frac{d\vec{p}}{(2\pi)^2}\Gamma^{(i)}_{\vec{V}_{1}\vec{V}}(\vec{p'}_{1},s'_{1},\vec{p}_{1},s_{1},\vec{p'},s', \vec{p},s) \nn\\
  &&\times G^{\vec{V}}_{s}(\vec{p},i\epsilon)G^{\vec{V}}_{s'}(\vec{p'},i\epsilon+i\omega)
  \Gamma^{(kj)}_{\vec{V}\vec{V}_{2}}(\vec{p},s,\vec{p'},s',\vec{p'}_{2},s'_{2}, \vec{p}_{2},s_{2};i\omega),\nn\\
  \label{new-vertex1.eq} 
  \eearr
  
  where $\vec{p'}=\vec{p}+\vec{q}$. The Green functions for eigenstates of the Hamiltonians Eq.~\ref{nonintH.eq} at
  $\mu=0$ have a simplest form, 
  
  \be
  G^{\pm \vec{K}}_{-}(\vec{p}, i\epsilon)=\frac{1}{i\epsilon-p},~~~~~ G^{\pm \vec{K}}_{+}(\vec{p}, i\epsilon)=\frac{1}{i\epsilon+p}
  \label{greenf-ne.eq}
  \ee
  
  By further processing, I should note that the vertices Eq.~\ref{vertices4.eq} can be decoupled in to $left$ and $right$ parts,
  
  \be
  \Gamma^{(i)}_{\vec{V}_{1}\vec{V}_{2}}(\vec{p'}_{1},s'_{1},\vec{p}_{1},s_{1},\vec{p'}_{2},s'_{2}, \vec{p}_{2},s_{2})=
  \Gamma^{i.\vec{V}_{1}}_{\vec{p'}_{1}s'_{1}\vec{p}_{1}s_{1}} . u .  \Gamma^{i.\vec{V}_{2}}_{\vec{p'}_{2}s'_{2}\vec{p}_{2}s_{2}},
  \label{l-r-gamma.eq}
  \ee
  
  where 
  
  \be
  \gamma^{a}_{\vec{p'}s'\vec{p}s}=\frac{1}{2}e^{\frac{i}{2}(-\phi+\phi')}, ~~~~~ \gamma^{b}_{\vec{p'}s'\vec{p}s}=
  \frac{1}{2}ss' e^{\frac{i}{2}(\phi-\phi')},
  \label{new-gamma.eq}
  \ee
  
  here the brief notations $1.\vec{K}=2.(-\vec{K})=a$ and $2.\vec{K}=1.(-\vec{K})=b$ are introduced.  Since all bare vertices are decoupled,
   the summarized vertices also can  be decoupled into the left and right parts. The RPA-renormalized interaction
   is written, $\Gamma^{ij}_{\vec{V}_{1}\vec{V}_{2}}=\Gamma^{i.\vec{V}_{1}}V_{ij}\Gamma^{i.\vec{V}_{2}}$. By substituting
   it into Eq.~\ref{new-vertex1.eq}, the interaction equations for this system obtained as,
   
   \be
   V_{ij}=u+u\sum_{k\vec{V}}\Pi^{\vec{V}}_{ik}V_{kj},
   \label{int-term.eq}
   \ee
  
  the polarization operators were introduced as,
  
  \bearr
  &&\Pi^{\vec{V}}_{ij}(\vec{q},i\omega)=\nn\\
  &&-T\sum_{\epsilon ss'}\int\frac{d\vec{p}}{(2\pi)2}\Gamma^{i.\vec{V}}_{\vec{p'}s'\vec{p}s}G^{\vec{V}}_{s}(\vec{p},i\epsilon)
  \Gamma^{i.\vec{V}}_{\vec{p}s\vec{p'}s'}G^{\vec{V}}_{s'}(\vec{p'},i\epsilon+i\omega).\nn\\
  \label{pol.eq}
  \eearr

    A nontrivial solution for the Eq.~\ref{int-term.eq} occurs when it's determinant is zero,
    
  \bearr
 \begin{vmatrix} 
 1-u\Pi^{\vec{K}}_{11}-u\Pi^{-\vec{K}}_{11} &    -u\Pi^{\vec{K}}_{12}-u\Pi^{-\vec{K}}_{12}\\ 
  -u\Pi^{\vec{K}}_{21}-u\Pi^{-\vec{K}}_{21}  &   1-u\Pi^{\vec{K}}_{22}-u\Pi^{-\vec{K}}_{22}
  \end{vmatrix}=\nn\\
  1-u(\Pi^{\vec{K}}_{11}+\Pi^{-\vec{K}}_{11}+\Pi^{\vec{K}}_{22}+\Pi^{-\vec{K}}_{22})\nn\\
  +u^{2}(\Pi^{\vec{K}}_{11}+\Pi^{-\vec{K}}_{11})(\Pi^{\vec{K}}_{22}+\Pi^{-\vec{K}}_{22})\nn\\
  -u^{2}(\Pi^{\vec{K}}_{12}+\Pi^{-\vec{K}}_{12})(\Pi^{\vec{K}}_{21}+\Pi^{-\vec{K}}_{21})=0,
  \label{determin3.eq}
 \eearr
  
   By using Eqs.~\ref{greenf-ne.eq},~\ref{l-r-gamma.eq},~\ref{new-gamma.eq} and Eq.~\ref{pol.eq}, the polarization operators can be
   easily calculated, by performing summation on $\epsilon=\pi T (2n+1)$ and on $s,s'=\pm1$, we find
   
   \bearr
   \Pi^{\vec{K}}_{11}=\Pi^{-\vec{K}}_{11}=\Pi^{\vec{K}}_{22}=\Pi^{-\vec{K}}_{22}=\frac{1}{2}\int\frac{d\vec{p}}{(2\pi)^2}\frac{p+p'}{(i\omega)^2-(p+p')^2},\nn\\
   \Pi^{\vec{K}}_{12}= \Pi^{-\vec{K}}_{21}=-\frac{1}{2}\int\frac{d\vec{p}}{(2\pi)^2}\frac{e^{i(-\phi+\phi')}(p+p')}{(i\omega)^2-(p+p')^2},\nn\\
   \Pi^{\vec{K}}_{21}= \Pi^{-\vec{K}}_{12}=-\frac{1}{2}\int\frac{d\vec{p}}{(2\pi)^2}\frac{e^{i(\phi-\phi')}(p+p')}{(i\omega)^2-(p+p')^2}.\nn\\
   \label{pol2.eq}
   \eearr
   
     Comparing Eq.~\ref{pol2.eq} with~\ref{chi-sub-lattices.eq} and~\ref{determin1.eq} with~\ref{determin3.eq}, we find that 
   $\chi^{0}_{AA}=\chi^{0}_{BB}=\Pi^{\vec{K}}_{11}+\Pi^{-\vec{K}}_{11}=\Pi^{\vec{K}}_{22}+\Pi^{-\vec{K}}_{22}$ and
   $\chi^{0}_{AB}=\chi^{0}_{BA}=\Pi^{\vec{K}}_{12}+\Pi^{-\vec{K}}_{12}=\Pi^{\vec{K}}_{21}+\Pi^{-\vec{K}}_{21}$. Therefore,
   the  sub-lattice representation give the same result as that in the representation of eigenstates in short range interaction.
   So in this representation we could not find zeros for Eq.~\ref{determin3.eq}. Finally we found that
   the on-site Hubbard interaction can not lead to formation of neutral collective mode. The essential reason is overlap factors
   which resolve sub-lattices. In the first approach the overlap factors reside in the Green functions, and in the second
   approach these factors are prescribed to interaction vertices.
    
    \section{Long range interaction: effective Coulomb interaction}
    
       The Coulomb interaction between electrons in graphene is usually considered as long-range, so both sub-lattices
     participate equally in it, since characteristic range of the interaction is much larger than the lattice constant.
     In this section we consider long range interaction for finding it's effect on spin-1 collective mode, which was considered
     in~\cite{JPHCM.Jafari}. If the long range (Coulomb) interaction is included interaction between electrons with same and 
     opposite spins on two sub-lattices, we know,  it results the plasmon collective mode~\cite{Wunsch}. The only long range interaction
     which lead to the Eq.~\ref{main.eq} has the form of Eq.~\ref{long-H-exch.eq}. Another reason for selecting this form for long range interaction is, the spin-1 collective
     mode make from interaction between electrons with opposite spins.

     If we assume that the Coulomb interaction is represented by  constant $V$ between electrons with opposite spin (because the spin-1 mode
     considered between two electrons with opposite spin), we introduced this interaction Hamiltonian,
     
     \bearr
     H_{Coul}=V\sum_{ij}\left \{a^{\uparrow \dagger}_{i}a^{\uparrow}_{i}a^{\downarrow \dagger}_{j}a^{\downarrow }_{j}+
     a^{\uparrow \dagger}_{i}a^{\uparrow}_{i}b^{\downarrow \dagger}_{j}b^{\downarrow }_{j}\right. \nn \\
     \left. +b^{\uparrow \dagger}_{i}b^{\uparrow}_{i}a^{\downarrow \dagger}_{j}a^{\downarrow }_{j} + 
     b^{\uparrow \dagger}_{i}b^{\uparrow}_{i}b^{\downarrow \dagger}_{j}b^{\downarrow }_{j}  \right \}.
     \label{long-H.eq}
     \eearr
   
   In the exchange channel, we have
   
   \bearr
     H_{Coul}=-V\sum_{ij}\left \{a^{\uparrow \dagger}_{i}a^{\downarrow}_{j}a^{\downarrow \dagger}_{j}a^{\uparrow }_{i}+
     a^{\uparrow \dagger}_{i}b^{\downarrow}_{j}b^{\downarrow \dagger}_{j}a^{\uparrow }_{i}\right. \nn \\
     \left. +b^{\uparrow \dagger}_{i}a^{\downarrow}_{j}a^{\downarrow \dagger}_{j}b^{\uparrow }_{i} + 
     b^{\uparrow \dagger}_{i}b^{\downarrow}_{j}b^{\downarrow \dagger}_{j}b^{\uparrow }_{i}  \right \}.
     \label{long-H-exch.eq}
     \eearr
   
    If we perform the same calculations starting from Eq.~\ref{H_int_new.eq}, as was done starting from Eq.~\ref{H_int_new.eq}
    to derive Eq.~\ref{determin3.eq}, we will arrive to the different system,

    \be
   \begin{cases}
    \chi_{AA}=\chi^{0}_{AA}+V\chi^{0}_{AA}\chi_{AA}+V\chi^{0}_{AB}\chi_{BA}+V\chi^{0}_{AA}\chi_{BA}+V\chi^{0}_{AB}\chi_{BA}\\
    \chi_{AB}=\chi^{0}_{AB}+V\chi^{0}_{AA}\chi_{AB}+V\chi^{0}_{AB}\chi_{BB}+V\chi^{0}_{AA}\chi_{BB}+V\chi^{0}_{AB}\chi_{BB}\\
    \chi_{BA}=\chi^{0}_{BA}+V\chi^{0}_{BA}\chi_{AA}+V\chi^{0}_{BB}\chi_{BA}+V\chi^{0}_{BA}\chi_{BA}+V\chi^{0}_{BB}\chi_{BA}\\
    \chi_{BB}=\chi^{0}_{BB}+V\chi^{0}_{BA}\chi_{AB}+V\chi^{0}_{BB}\chi_{BB}+V\chi^{0}_{BA}\chi_{BB}+V\chi^{0}_{BB}\chi_{BB}\\
  \end{cases}.
     \label{coupling2.eq}
  \ee

    with the solvable condition 
    
    \be
    1-V(\chi^{0}_{AA}+\chi^{0}_{AB}+\chi^{0}_{BA}+\chi^{0}_{BB})=0,
    \label{main.eq}
    \ee
    
    where the susceptibilities are same as Eq.~\ref{chi-sub-lattices.eq}. This equation is same as with essential 
    equation of~\cite{Ebrahimkhas-IJMPB} which predicted neutral collective mode in honey comb lattice. 
    If we consider $ \chi^{0}_{triplet}=\chi^{0}_{AA}+\chi^{0}_{AB}+\chi^{0}_{BA}+\chi^{0}_{BB}$, the dispersion of
    the neutral collective mode can be obtained with one of this equations,
    
    \be
    \Re{\chi^{0}_{triplet}(\vec{q},\omega)}=1/V~~~~or~~~~~ \Im{\chi^{0}_{triplet}(\vec{q},\omega)}=0.
    \label{condition.eq}
    \ee

\begin{figure}[h]
\begin{center}
\vspace{-3mm}
\includegraphics[width=6cm,height=5cm,angle=0]{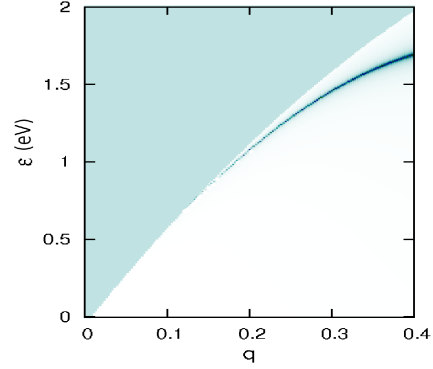}
\vspace{-3mm}
\caption{The contour plot of $\Re{\chi^{0}_{triplet}(\vec{q},\omega)}=1/V$ is displayed 
for undoped graphene for V=3.2t. The dispersion of spin collective mode appeared below e-h continuum (cyan region) but not in
reasonable range of $V$.
The $y$ axis is energy ($\epsilon (eV)$)  $vs$  momentum ($q$). The direction in e-h continuum is  $\Gamma \rightarrow K$.}
\label{e-h-dis.fig}
\end{center}
\vspace{-3mm}
   \end{figure}

     The zeros of the Eqs.~\ref{condition.eq} are  collections of $(\vec{q}, \omega)$ which
     are plotted in Fig.~\ref{e-h-dis.fig} for $V=3.2t$. This contour plot displays dispersion of neutral collective mode.
     The contour plot of $\Re \chi^{0}_{triplet}$ for $V=3.2t$ is plotted under the electron-hole $(e-h)$ continuum 
     region in $\Gamma \rightarrow K$ direction. But we know this range of $U$ is too large for electrons in graphene.
     I checked the solution and zeros of Eq.~\ref{condition.eq} for $0.0t<V<3.0t$, but I didn't find any zeros. For $V>3.2t$
     we could find dispersions for neutral collective mode near $\Gamma$ point in $\Gamma-K$ direction which are meaningless.
     So  the long range interaction of type Eq.~\ref{long-H.eq} can not formed the neutral collective mode
    in undoped graphene without $one-cone approximation$.
     
        The prediction of triplet collective mode in Ref.~\cite{Ebrahimkhas-IJMPB} was due to
      the $single-cone$ approximation in short range  on-site interaction that led to scalar form of polarization operator
      without $U^2$ terms in Eq.~\ref{determin2.eq}. One-cone approximation 
      which lead to elimination of some parts of Eq.~\ref{determin2.eq}, is the only reason that predicted of the spin-1 collective mode.

    \section{Conclusion}

         In this research the effects of short range Hubbard interaction in two representations and new 
     effective long range interaction Hamiltonian was studied and introduced. The prediction of
     neutral collective mode in on-site and effective long range interaction is analyzed, and  we found that in the RPA method
     on Hubbard model without single-cone approximation, mixing the overlap functions of two sub-lattices suppressed
     this collective mode. These two types of interactions have
     different Hamiltonians, Eq.~\ref{H_int_new.eq} and Eq.~\ref{long-H-exch.eq}, in both I considered interaction between
     electrons from the electrons  with opposite spins. We can't find and  predict neutral collective mode as the effect of
     effective long range  and short range interaction because the overlap factors in both case resolve sub-lattices and suppress 
     the neutral collective mode.

  \section{Acknowledgment}
  
    I wish to thank N. Sharifzadeh for helpful discussions.
    The author appreciates and thanks so much from referee of our paper~\cite{Ebrahimkhas-IJMPB} who guides to reach the main idea of this
    paper.

\end{document}